\documentclass[a4paper,10pt]{article}

\usepackage[toc,page]{appendix}

\usepackage{ae}

\usepackage[ansinew]{inputenc}

\usepackage{lscape}

\usepackage{amssymb,amsmath,amsthm}

\usepackage{epsf}

\usepackage{psfrag}

\usepackage{mathrsfs}

\theoremstyle{definition}

\newtheorem{thm}{Theorem}

\newtheorem{Remark}{Remark}

\newtheorem{Conjecture}{Conjecture}

\newcommand{\metric}{{\bf g}}

\newcommand{\myK}{K}

\title{The spherically symmetric Einstein-scalar field system with positive and vanishing cosmological constant: a comparison.}

\author{Jo\~ao L. Costa \\
{\small Instituto Universitário de Lisboa (ISCTE-IUL), Lisboa, Portugal}\\
{\small Centro de An\'alise Matem\'atica, Geometria e Sistemas Din\^amicos,}\\
{\small Instituto Superior T\'ecnico, Universidade T\'ecnica de Lisboa, Portugal}
}

\begin{document}

\maketitle

\begin{abstract}

We review recent results concerning the spherically symmetric  Einstein-scalar field system with positive cosmological constant. We do so by  comparing with the classical results of Christodoulou concerning the asymptotically flat case (vanishing cosmological constant) and by discussing some of the issues which have emerged since the publication of our main results.

Concerning the positive cosmological constant case, we also sketch an alternative proof of global in (Bondi) time existence, based on energy estimates, which is presumably more flexible and, consequently, amenable to generalizations; other potential improvements and generalizations of our main results are also discussed.

%
\end{abstract}

\section{Introduction}

 Following Hilbert's advice, concerning the importance of specialization, Christodoulou was led to the spherically symmetric Einstein-scalar field system as a suitable model problem for the study of gravitational collapse: it gives rise to a non-trivial (as opposed to Einstein-Maxwell), non-pathological (as opposed to Einstein-dust), $1+1$ system of partial differential equations that moreover shares the wave character of the
 general Einstein-vacuum equations. This was the starting point of Christodoulou's tour de force concerning self-gravitating scalar fields, that led to extraordinary insights concerning gravitational collapse, culminating in the celebrated formation of trapped surfaces Theorem~\cite{Christodoulou:2008}. Here we will be focusing on the first and simplest of the steps of this quest, which concern the problem of dissipation for small data~\cite{Christodoulou:1986}, and mostly to its recent extension with the inclusion of a positive cosmological constant~\cite{CostaProblem}; note that the previous motivations remain valid in the presence of a non-vanishing cosmological constant~\footnote{In particular, for a negative cosmological constant~\cite{HolzegelSelf,HolzegelStability}, which we will not discuss here.}.


From a physical point of view, introducing a positive cosmological constant into the Einstein field equations  provides the simplest known mechanism to model inflation periods (large $\Lambda$), as well as the ``recent" period of accelerated expansion (small $\Lambda$), and consequently plays a central role in modern cosmology.
From a purely mathematical point of view, such introduction leads to a all new range of dynamical behaviors and geometrical structures.
All this adds to the relevance of studying initial value problems for the Einstein-matter field equations with positive cosmological constant. For such problems a general framework is provided by the following conjecture:
\begin{Conjecture}
{(Cosmic ``no-hair" conjecture)} { Generic expanding solutions of Einstein's field equations with a positive cosmological constant approach the de Sitter solution asymptotically}.
\end{Conjecture}
As usual, part of the challenge of the conjecture is to obtain a precise statement to it. This conjecture has been verified for a variety of matter models and/or symmetry conditions (see~\cite{CostaProblem} for a brief overview), but the complexity of the issue makes a general result unattainable in the near future. For instance, either by symmetry conditions or smallness assumptions on the initial data the formation of cosmological black holes is excluded from all known realizations of this statement.

The main goal of the present paper is to review the recent results concerning the spherically symmetric  Einstein-scalar field system with positive cosmological constant~\cite{CostaProblem}. This will be done by a comparison with the classical results of Christodoulou concerning the asymptotically flat case (vanishing cosmological constant) and by discussing some of the issues which have emerged since the publication of our main results.

We start by showing how the setup originally developed for the asymptotically flat case naturally accommodates the introduction of a cosmological constant. Then the  strategy used in  the classic asymptotically flat case~\cite{Christodoulou:1986} is reviewed while providing an explanation of some of the basic nonlinear analysis  techniques needed; we have tried to give special emphasis to the points which provide a clearer comparison of the differences between the $\Lambda=0$ and $\Lambda>0$ cases.  We then discuss how the introduction of a positive cosmological constant requires a considerable deviation from the original strategy, explain the main ideas of the proof of global existence and exponential decay in Bondi time, developed in~\cite{CostaProblem}, and discuss such results in the context of the cosmic ``no-hair" conjecture.

Also concerning the cosmological case, we  provide ideas for an alternative proof of global in (Bondi) time existence, based on energy estimates, which is presumably more flexible and, consequently, amenable to generalizations which might, for instance, include some classes of nonlinear scalar fields.  We also discuss other potential improvements of the main results presented, as well as future plans of research, including large data analysis of the system under consideration.

\section{Setup and main results}
\label{sectionSetup}

\subsection{Bondi spherical symmetry}

A spacetime $(M, \metric)$ is {\em Bondi-spherically symmetric} if it admits a global representation for the metric of the form
\begin{equation}
\label{metricBondi}
 \metric=-f(u,r)\tilde{f}(u,r)du^{2}-2f(u,r)dudr+r^{2}\sigma_{\mathbb{S}^2}\;,
\end{equation}
with: $\sigma_{\mathbb{S}^2}$
the round metric of the two-sphere; the {\em radius function} defined by $r(p):=\sqrt{\text{Area}({\mathcal O}_p)/4\pi}$, where ${\mathcal O}_p$ is the orbit of an $SO(3)$ action by isometries through $p$;  the future null cones of points at $r=0$  given by $u=constant$; $u$ is also known as the Bondi time.

 For instance, the causal future of any point in de Sitter spacetime may be covered by Bondi coordinates with the metric given by
\begin{equation}
 \mathring{\metric}=-\left(1-\frac{\Lambda}{3}r^{2}\right)du^{2}-2dudr+r^{2}\sigma_{\mathbb{S}^2}\;.
\label{dSBondi}
\end{equation}
Contrary to what happens with Minkowski spacetime, in the $\Lambda=0$ case, this coordinate system does not cover the full de Sitter manifold, which strictly speaking is not Bondi-spherically symmetric. Nonetheless Bondi coordinates cover the region of interest to us here: the full asymptotic region and in particular a neighborhood of any proper subset of $\mathscr{I^+}$.

\begin{figure}[h!]

\begin{center}

\psfrag{u=+i}{$u=+\infty$}

\psfrag{u=-i}{$u=-\infty$}

\psfrag{u=0}{$u=0$}

\psfrag{H}{$\mathcal{H}$}

\psfrag{H'}{$\mathcal{H'}$}

\psfrag{I+}{$\mathscr{I^+}$}

\epsfxsize=.5\textwidth

\leavevmode

\epsfbox{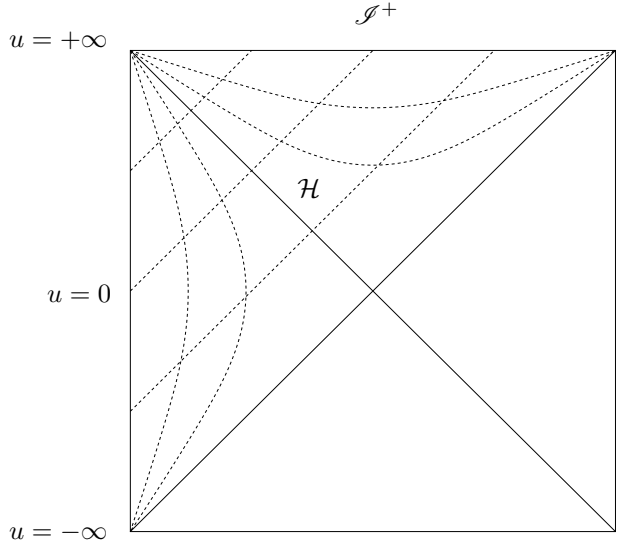}

\end{center}

\caption{Penrose diagram of de Sitter spacetime. The dashed lines $u=\text{constant}$ are the future null cones of points at $r=0$. The cosmological horizon $\mathcal{H}$ corresponds to $r=\sqrt{\frac3\Lambda}$ and future infinity $\mathscr{I^+}$ to $r=+\infty$.} \label{Penrose}

\end{figure}

Another drawback of the Bondi ansatz is that it excludes the Nariai solution, for which the radius function is everywhere constant and  cannot be used as a coordinate. Such spherically symmetric solution of the Einstein vacuum equation with positive cosmological constant is especially relevant since it is expected to provide a counterexample for a naive version of cosmic no-hair where no genericity assumptions are imposed~\cite{Beyer:2010}. For us here, the a priori exclusion of Nariai is not problematic since we will be focused on nonlinear perturbations of de Sitter spacetime.

The use of double null coordinates would allow us to bypass these drawbacks of the Bondi setup. It would also potentially simplify the  analysis to come at some points: for instance, the characteristics of the problem have a trivial description in such framework  (compare with Section~\ref{sectionCharact}). Also, since they have become the standard coordinate system in the mathematical analysis of spherically symmetric gravitational collapse, its use might be beneficial in the name of homogeneity within the literature on the subject.

So what are the advantages of Bondi coordinates and why use them?
First of all, in double null coordinates, the Einstein equations reduce to a system of partial differential equations which is singular at the center of symmetry $r=0$. For this reason local well posedness for such system usually requires $r$ to be bounded away from zero.
On the other hand, the Bondi ansatz allows us to reduce the full content of the Einstein equations to a single scalar integro-differential equation~\eqref{mainEq}, which turns out to facilitate the handling of the center of symmetry; this has a quite remarkable manifestation  in the natural way by which regularity at the axis is obtained. One should also note that the cornerstones of the analysis reviewed here were discovered and are more naturally formulated within the Bondi framework.


\subsection{Main result}

The main results in~\cite{CostaProblem} establish well-posedness, global existence  and exponential decay in (Bondi) time of the Einstein-scalar field system with $\Lambda>0$, for small spherically symmetric data given in a truncated null cone. It follows that initial data close enough to de Sitter data evolves to a causally geodesically complete spacetime (with boundary), which approaches a region of de Sitter asymptotically at an exponential rate; a realization of the cosmic no-hair conjecture. More precisely we have:

\begin{thm}
\label{mainThm}

Let $\Lambda>0$ and $R>\sqrt{3/\Lambda}$. There exists $\epsilon_0>0$, depending on $\Lambda$ and $R$, such that for  $\phi_0\in{\mathcal C}^{k+1}([0,R])$, $k \geq 1$, satisfying
$$\sup_{0\leq r\leq R }|\phi_0(r)|+\sup_{0\leq r\leq R }|\partial_r\phi_0(r)|<\epsilon_0\;,$$
there exists a unique Bondi-spherically symmetric ${\mathcal C}^k$ solution
$(M,\metric,\phi)$ of the Einstein-$\Lambda$-scalar field system
\begin{equation}
\label{fieldEq}
 R_{\mu\nu}=\kappa\, \partial_{\mu}\phi\,\partial_{\nu}\phi+\Lambda \bf{g}_{\mu\nu}\;,
\end{equation}
with the scalar field $\phi$ satisfying the characteristic condition
$$\phi_{|_{u=0}}=\phi_0\;.$$
The Bondi coordinates for $M$ have range
$[0,+\infty)\times[0,R]\times \mathbb{S}^2$,
and the metric takes the form~\eqref{metricBondi}. Moreover, we have the following bound in terms of initial data:
$$
\left|\phi\right| \leq \sup_{0\leq r\leq R }\left|\partial_r\left(r\phi_0(r)\right)\right|\;.
$$
Regarding the asymptotics, there exists $\underline{\phi}\in\mathbb{R}$ such that
\[
\label{phiDecay}
\left|\phi(u,r)-\underline{\phi}\right|\lesssim e^{-2Hu}\;,
\]
and
\[
\label{metricDecay}
\left|\metric_{\mu\nu}-\mathring{\metric}_{\mu\nu} \right|\lesssim e^{-2Hu}\;,
\]
where $H:=\sqrt{\Lambda/3}$ and $\mathring{\metric}$ is de Sitter's metric in Bondi coordinates, as given in~\eqref{dSBondi}. Finally, the spacetime $(M,\metric)$ is causally geodesically complete towards the future and has vanishing final Bondi mass.
\end{thm}

Some comments are in order:

First note that solutions whose scalar fields  differ by a constant have the same physical content since the field equations~\eqref{fieldEq} ``do not see constants". In particular this means that we can always perform a rescaling of the data that results in the vanishing of the asymptotic constant $\underline{\phi}\in\mathbb{R}$ without changing the geometric and dynamical content of the solution. Note as well that the linear theory~\cite{Rendall:2004} tells us to expect $\phi$ to converge, when $r\rightarrow\infty$, to a (not necessarily constant) function $\phi_{\infty}(u)$; the same behavior is expected in the nonlinear case once a solution covering the entire radial range is obtained; remarkably such expectation does not necessarily contradict cosmic ``no hair" if for instance $\metric(\nabla \phi,\nabla \phi)$ converges to zero in an appropriate sense.

Since our solutions are only defined up to $r=R$ the spacetime manifolds they provide   are  manifolds with boundary. By geodesic completeness towards the future we then mean that the only geodesics which cannot be continued for all values of the affine parameter are those with endpoints on the boundary $r=R$. Also, from the presence of such boundary one might expect the necessity to impose boundary conditions, at $r=R$, as well. A more detailed justification on why this isn't necessary has to be postponed to Section~\ref{SectionIteration}, but let us just say for now that the fact that one can get a global in (Bondi) time result from local (truncated) initial data is a recurrent feature of cosmological solutions undergoing accelerated expansion, see~\cite{Ringstrom:2008}. Note as well that although we can find solutions for arbitrarily large $R$, the upper bound on the size of the initial data $\epsilon_0\rightarrow 0$, as $R\rightarrow\infty$.

We clearly do not expect the solution obtained to correspond to the maximal globally hyperbolic development of the data considered. The reason for such undesirable incompleteness originates in the (undesirable) necessity to impose $r\leq R$ in the full domain and not only at the level of initial data. It turns out that such restriction is not directly related to causal issues but to the need of controlling quantities that grow exponentially with $r$. Such issues will be discussed in detail throughout.

Again in view of the restriction on the radial range, the usual definition of Bondi mass has to be abandoned, since it involves a limit at $r=\infty$. The Bondi mass referred to in the theorem above was introduced in~\cite{CostaProblem} and corresponds to a mass taken along the apparent cosmological horizon; see Section~\ref{SectionMass} for more details.

Another source of confusion comes from the fact that one might expect regularity at the center to necessarily require
$$\partial_r\phi(u,0)=0\;.$$
But this was not even imposed at the level of initial data and, in fact, besides being of class $\mathcal{C}^k$, initial data has no additional constraints!
This is a manifestation of the famous {\em fundamental confusion of calculus}. The correct regularity condition is in fact
$$\partial_r\phi(u,0)-\partial_u\phi(u,0)=0\;,$$
i.e., the vanishing of the derivative, at $r=0$, along vectors orthogonal to the center of symmetry; it turns out that such condition follows from the wave equation! Meaning that we can freely specify the data, solve the equation and obtain the desired regularity for free.
As an instructive example note that the smooth function $\phi(t,r)=t$ is
the solution of the spherically symmetric  wave equation in Minkowski, with initial condition $\phi(r,r)=r\;.$

We finish this preliminary discussion with some comments concerning the exponential qualification of the decay: note that, in de Sitter,  our retarded time coordinate $u$ in~\eqref{dSBondi} and the standard (flat FLRW) time coordinate $t$ are related by
%
%
%
\begin{equation}
\label{coordRelation}
u=t-\sqrt{3/\Lambda}\,\log\left(1+\sqrt{\Lambda/3}\,r\right)\;.
\end{equation}
So we see that, for $r\leq R$, exponential decay in $u$ corresponds to exponential decay in $t$ in those standard coordinates.

Let us return to the de Sitter perturbations provided by our main result: although in the Bondi framework there is a gauge freedom of the form $u\mapsto \psi(u)$, which might render the classification of the decay as exponential meaningless without an explicit characterization of the gauge fixing, it turns out that since the metric components decay exponentially in $u$ to de Sitter's metric components~\eqref{dSBondi} the desired qualification of the decay  follows in view of~\eqref{coordRelation}. We take the chance to note that we have used  the gauge fixing condition $f(u,r=0)\equiv 1$, which corresponds to setting the clock at the center of symmetry.   Christodoulou's choice is $f(u,r=\infty)\equiv 1$, a choice  unavailable to us in view of the finite radial range considered. For an asymptotic analysis fixing the gauge at infinity seems more natural: first because it corresponds to a partial standard description of asymptotical flatness in  $\Lambda=0$ and of being asymptotically de Sitter in the $\Lambda>0$ case; secondly, by setting the clock at the center then, in a potential solution covering the full radial range, one expects $f$  to converge to a (not necessarily constant) function $f_{\infty}(u)\geq 1$, as $r\rightarrow\infty$, making the qualification of the decay, in $r$, away from $u=\infty$ less clear.


One should also note that the linear analysis in~\cite{Rendall:2004} gives rise to the decay $e^{-Ht}$ while we are obtaining $e^{-2Hu}$; the extra decay is presumably a consequence of the fact that $\frac{\partial}{\partial u}$ becomes tangent to the cosmological horizon, as $u$ becomes large, and as a consequence red-shift is added to the damping effect of the accelerated expansion.

\subsection{The Einstein-$\Lambda$-scalar field in Bondi coordinates.}

Let $G_{\mu\nu}$ denote the Einstein tensor. In Bondi-spherical symmetry, the full content of the field equations is encoded  in the Einstein equations:
\begin{equation}
G_{rr}=0\Leftrightarrow \frac{2}{r}\frac{1}{f}\frac{\partial f}{\partial r}=\kappa\left(\partial_{r}\phi\right)^{2}\;,
\end{equation}
\begin{equation}
G_{\theta\theta}=0\Leftrightarrow \frac{\partial}{\partial r}(r\tilde{f})=f\left(1-\Lambda r^{2}\right)\;,
\end{equation}
and the wave equation for the scalar field,
$$
\nabla^{\mu}T_{\mu\nu}=0 \Leftrightarrow \nabla^{\mu}\partial_{\mu}\phi=0\;,
$$
which reads
\begin{equation}
\frac{1}{r}\left[\frac{\partial}{\partial u}-\frac{\tilde{f}}{2}\frac{\partial}{\partial r}\right]\frac{\partial}{\partial r}\left(r\phi\right)=\frac{1}{2}\left(\frac{\partial\tilde{f}}{\partial r}\right)\left(\frac{\partial\phi}{\partial r}\right)\;.
 \end{equation}

We have already set up the wave equation in order to introduce the change of variable
$$h={\partial}_r(r\phi)\;.$$
The motivation for this comes from the fact that, in  Minkowski, the  spherically symmetric wave equation becomes
\begin{equation}
\label{waveMinkowski}
\quad\quad Dh=0 \quad,\quad D=\partial_u-\frac{1}{2}\partial_r\;,
\end{equation}
a trivial transport equation! If we introduce a cosmological constant, this change of variable leads to the following form for the wave equation:
\begin{equation}
\label{wavedeSitter}
Dh=-\frac{\Lambda}{3}r(h-\bar{h}) \quad,\quad D=\frac{\partial}{\partial u}-\frac{1}{2}\left(1-\frac{\Lambda}{3}r^{2}\right)\frac{\partial}{\partial r}\;.
\end{equation}
Clearly the simplification for $\Lambda\neq 0$ is not as striking. Although we do not know weather the introduction of $h$ is optimal in this case,  one thing is clear: for $\Lambda>0$, it cannot be expected that another change of variables will lead to the dramatic simplification that occurred in the asymptotically flat case, since, in de Sitter, solutions to the wave equation decay exponentially in time~\cite{Rendall:2004,CostaSpherical}.

Returning to the nonlinear setting, applying the previously motivated change of variable encodes the full content of the field equations~\eqref{fieldEq} in the integro-differential scalar equation:
\begin{equation}
\label{mainEq}
Dh=G\left(h-\bar{h}\right)\;,
\end{equation}
where
\begin{equation}
\label{D}
 D=\frac{\partial}{\partial u}-\frac{\tilde{f}}{2}\frac{\partial}{\partial r}
\quad\quad \text{ , } \quad\quad G=\frac{1}{2r}\left[(f-\tilde{f})-\Lambda f r^{2}\right]\;,
\end{equation}
and with the radial--average of a function given by
\begin{equation}
\label{scalar}
{\bar h}:=\frac{1}{r}\int_0^rh(u,s)ds\;.
\end{equation}
It turns out that the scalar field is recovered from $h$ by averaging:
\begin{equation}
\label{scalar}
\phi={\bar h}\;.
\end{equation}
As for the remaining unknowns, setting $f(u,r=0)=1$, which can always be done by an appropriate rescalling $u\mapsto \psi(u) $, the metric coefficients are obtained from $h$ by the relations:
\begin{equation}
\label{metricQuoficients}
f(u,r)=\exp\left({\frac{\kappa}{2}\int^{r}_{0}\frac{\left(h-\bar{h}\right)^{2}}{s}ds}\right) \quad,\quad\tilde{f}(u,r)=\bar{f}-\frac{\Lambda}{r}\int^{r}_{0}fs^{2}ds\;.
\end{equation}

We will also need an evolution equation for $\partial_rh$ given a sufficiently regular solution of~\eqref{mainEq}; this turns out to be
\begin{equation}
\label{D_partial_h}
 D\partial_{r}h-2G\partial_{r}h=-J\,\partial_{r}\bar{h}\;,
\end{equation}
for
\begin{equation}
\label{defJ}
J:=3G+\Lambda g r+(\Lambda r^{2}-1)\frac{1}{2}\frac{\partial g}{\partial r}\;.
\end{equation}

\section{Christodoulou and the asymptotically flat case, $\Lambda=0$.}
\label{SectionLambda0}

In~\cite{Christodoulou:1986}, although not stated as such, a result along the lines of Theorem~\ref{mainThm} was established for the asymptotically flat case, $\Lambda=0$. The full radial range $R=+\infty$ is considered and, concerning asymptotics, it is shown that, for appropriately decaying initial data, $\phi=\bar h$, $h$ and $\partial_rh$ all decay to zero at a polynomial rate in both $u$ and $r$. As a consequence, the corresponding spacetime is future geodesically complete and approaches Minkowski asymptotically.

To appreciate the differences  between the vanishing and the positive cosmological constant cases we give a brief overview of the strategy developed by Christodoulou to attack the asymptotically flat case. Hopefully this will be of interest on its own for those readers which are not familiarized with such work and underlying techniques. Later we will see how the introduction of a positive $\Lambda$ requires a considerable deviation from the original strategies.

After reducing the field equations to the integro differential equation~\eqref{mainEq}~\footnote{The setup described in section~\ref{sectionSetup} works for any value of $\Lambda$, and to recover the original asymptotically flat case one only has to set $\Lambda=0$. Nonetheless, recall that Christodoulou fixes his $u$-coordinate by demanding that $f(u,r=\infty)\equiv1$, i.e., the clock is set by the observers at infinity rather then the observers at the origin, as used in this paper, with the exception of the present section where the original gauge fixing is considered.},  Christodoulou considers the following norms:
for $h:[0,+\infty[\times[0,+\infty[\rightarrow \mathbb{R}$ define
$$\|h\|_{Y'}=\sup_{u,r\geq 0}\left\{\left(1+r+\frac{u}{2}\right)^3|h(u,r)|\right\}\;,$$
and
$$\|h\|_{X'}=\sup_{u,r\geq 0}\left\{\left(1+r+\frac{u}{2}\right)^3|h(u,r)|+\left(1+r+\frac{u}{2}\right)^4|\partial_rh(u,r)|\right\}\;,$$
where the reason for the $1/2$ factor in the decay in $u$ can be traced back to the (linear) wave equation in Minkowski~\eqref{waveMinkowski}, see also~\eqref{charMinkowski}.
Denote by $Y'$  the set of continuous functions with finite $\|\cdot\|_{Y'}$ norm and by $X'$ the set of continuous functions, with continuous partial $r$-derivative and finite  $\|\cdot\|_{X'}$ norm. Both $Y'$ and $X'$ are Banach spaces~\footnote{A Banach space is a complete normed vector space, complete meaning that every Cauchy sequence is convergent.}.  Clearly $X'\subset Y'$ and if $h\in X'$
\begin{equation}
\label{decayX'}
|h(u,r)|\leq \frac{\|h\|_{X'}}{(1+u/2+r)^3}\;\;\text{ and } \;\; |\partial_rh(u,r)|\leq \frac{\|h\|_{X'}}{(1+u/2+r)^4}\;.
\end{equation}

Now, given initial data $h(0,r)$ such that
%
$$\sup_{r\geq 0}\left\{(1+r)^3|h(0,r)|+(1+r)^4|\partial_rh(0,r)|\right\}=d<\infty\;,$$
let
$$h_0(u,r)=h\left(0,\frac{u}{2}+r\right)\;,$$
and note that $\|h_0\|_{X'}=d$.
With $h_0$ as seed we can now iteratively define a sequence by solving the linear equation
\begin{equation}
\label{iteration}
D_{n}h_{n+1}-G_{n}h_{n+1}=-G_{n}\bar{h}_{n}\;,
\end{equation}
with initial data $h_{n+1}(0,r)=h(0,r)$. The previous equation, which in essence is simply obtained by getting the nonlinear information from the previous iteration,  can be integrated to obtain
\begin{equation}
\label{iterationIntegral} h_{n+1}(u_1,r_1)=h_{0}(\chi_n(0))e^{\int^{u_1}_{0}G_{n|_{\chi_{n}}}dv}
-\int^{u_1}_{0}\left(G_{n}\bar{h}_{n}\right)_{|_{\chi_{n}}}e^{\int^{u_1}_{u}G_{n|_{\chi_{n}}}dv}du\;\;,
\end{equation}
where $\chi_n(u)=\chi_n(u;u_1,r_1)=(u,r_n(u;u_1,r_1))$ is the characteristic of the equation through $(u_1,r_1)$, i.e., the integral curve of the vector field $D_n=\partial_u-\frac{1}{2}\bar f_n\partial_ r$ that contains $(u_1,r_1)$; note that $\Lambda=0$ implies $\tilde f = \bar f$. The characteristics, which correspond to incoming null rays, satisfy the equation
$$\frac{dr_n}{du}=-\frac{1}{2}\bar f_n\;,$$
from which, using the upcoming~\eqref{fLambda0}, we obtain the following estimate in terms of Minkowskian characteristics:
\begin{equation}
\label{charMinkowski}
r_1+\frac{1}{2}K_0(u_1-u) \leq r_n(u)\leq r_1+\frac{1}{2}(u_1-u)\;.
\end{equation}

The goal is now to show that the sequence~\eqref{iterationIntegral} converges (or at least has a sublimit) in $X'$, given appropriately small initial data; the limiting function will then be seen to be a solution to the problem at hand.

First, one needs to show that, for small enough initial data (small enough $d$), there exists $x>0$, such that
\begin{equation}
\label{bola}
   \|h_n\|_{X'}<x, \text{ for all }n,
\end{equation}
with $x\rightarrow 0$, as $d\rightarrow 0$.
One proceeds by induction.
There is a considerable amount of estimating to be done and of course we will not be going through the full details; to exemplify  the kind of estimates one is interested in we present the most fundamental ones: starting from~\eqref{decayX'} and using the induction hypothesis $\|h_n\|_{X'}<x$ one obtains
%
\begin{equation}
\label{h-barhLambda0}
|h_{n+1}(u,r)-\bar h_{n+1}(u,r)|\leq C\; \frac{x}{(1+u/2)^2(1+u/2+r)}\;,
\end{equation}
\begin{equation}
\label{fLambda0}
e^{-Cx^2}=:K_0\leq f_{n+1} \leq 1 \equiv f_{Minkowski}\;,
\end{equation}
and
\begin{equation}
\label{GLambda0}
|G_{n+1}(u,r)|\leq C \frac{x^3\, r}{(1+u/2)^7(1+u/2+r)^4}\;,
\end{equation}
for some positive constants $C$ not depending on $n$.

First note that~\eqref{h-barhLambda0} also provides an estimate for the scalar field $\phi=\bar h$; an interesting aspect of such estimate is that for any choice of exponents of decay in the definition of the norm $X'$, $k$ for $h$ and $k+1$ for $\partial_r h$, with $k>1$, the scalar field will decay radially only has $r^{-1}$. Christodoulou's choice of $k=3$ seems then to be made for convenience, since it is the choice leading to the simplest computations.

Estimate~\eqref{fLambda0} is a cornerstone of the analysis: recall for instance that it allows one to  obtain estimates for the characteristics~\eqref{charMinkowski}. A important aspect of it, especially when comparing with the upcoming $\Lambda>0$ case, is  the fact that it is uniform in $r$.

Concerning the zeroth order term $G_{n+1}$, two aspects of estimate~\eqref{GLambda0} are of particular relevance now and for the future comparison: the fact that the  parameter $x$, controlling the size of the scalar field, appears cubed, and the fact that $G_{n+1}$ inherits its rate of radial decay from $h_n$.

Using, among others, the previous estimates, one can  control, via~\eqref{iterationIntegral}, the size of $(1+\frac{u}{2}+r)^3|h_{n+1}(u_1,r_1)|$. The same kind of ideas also allows us to control $(1+\frac{u}{2}+r)^4|\partial_r h_{n+1}(u_1,r_1)|$, this time by using the iteration scheme obtained from the upcoming~\eqref{drhn} by setting $\Lambda=0$; note that this in turn follows from an evolution equation for $\partial_r h_{n+1}$ obtained by differentiating~\eqref{iteration} with respect to $r$.  One can eventually conclude that
\begin{equation}
\label{bola2}
\|h_{n+1}\|_{X'}< C_1(1+x^2)e^{C_2x^2}(d+C_3x^3)\;.
\end{equation}
Appropriate choices of $d$ and $x$ then lead to the desired $\|h_{n+1}\|_{X'}\leq x$\;. One should note that the cubic term plays an important role here; in fact, it could be replaced by any power $x^p$, $p>1$, for the result to hold, but the linear case $p=1$ would only lead to the desired conclusion if luck would have the constants in the expression to satisfy $C_1C_3<1$.

Christodoulou then proceeds to show that the sequence $h_n$ contracts in $Y'$, that is
$$\|h_{n+1}-h_n\|_{Y'}\leq \sigma \|h_{n}-h_{n-1}\|_{Y'}\;,$$
for some $\sigma<1$. From this we see that $h_n$ is Cauchy and therefore convergent to some function $h\in Y'$. To see that this corresponds to a solution of the problem at hand with the desired regularity and decaying properties requires further work.

The next step is to see that $h$ is in fact in $X'$. To prove this Christodoulou uses the Arzel\`a-Ascoli Theorem which requires equicontinuity: Equicontinuity of $h_n$ follows from the equiboundedness of $\partial_rh_n$ provided by~\eqref{bola}; similarly, by using~\eqref{drhn}, we conclude that  $D_n\partial_rh_{n+1}$ is equibounded and
the equicontinuity of $\partial_r h_{n}$ with respect to $u$ follows.
To prove equicontinuity of $\partial_r h_n$ with respect to $r$ define
$$\psi(u)=\partial_r h_{n+1} \circ\chi_{n}(u;u_1,r_2)-\partial_r h_{n+1}\circ\chi_{n}(u;u_1,r_1)\;,$$
for given $u_1,r_1,r_2\geq0$. Differentiating with respect to $D_n$ and using~\eqref{drhn} gives an expression for $\psi'(u)$ which exclusively involves sequences which  are equicontinuous in $r$: in some cases one can see this directly and in others one can use~\eqref{bola} to conclude that the respective $r$ derivatives are equibounded. Then, given $\delta>0$, for appropriately small $|r_2-r_1|$, we have
$$\psi'(u)\leq C\delta\;,$$
uniformly for $u\leq u_1$ and with $C$ allowed to depend on $u_1$. Integrating the last inequality we are able to estimate the modulus of continuity, in $r$, of $\partial_rh_{n+1}$,  in terms of the respective modulus of continuity of the initial data and an arbitrarily small $\delta$; in particular, independently of $n$, and the desired equicontinuity follows.

We can now invoke the Arzel\`a-Ascoli Theorem to conclude that appropriate subsequences of $h_n$ and $\partial_r h_n$ converge uniformly, on (arbitrary) compact sets of the form $[0,U]\times[0,R]$, to functions $\tilde h$ and $\partial_r \tilde h$. Clearly  $\tilde h$ and $\partial_r \tilde h$ are defined and continuous in the full domain $[0,\infty)\times[0,\infty)$, and, relying once more on~\eqref{bola}, we conclude that, in fact, $\tilde h \in X'\subset Y'$. It follows that $h=\tilde h\in X'$\;.

By taking limits on both sides of~\eqref{iterationIntegral} $h$ appears as a solution of the integral equation
\begin{equation*}
h(u_1,r_1)=h_{0}(\chi(0))e^{\int^{u_1}_{0}G(\chi(v))dv}
-\int^{u_1}_{0}\left(G\bar{h}\right)\circ\chi(u)\,e^{\int^{u_1}_{u}G(\chi(v))dv}du\;\;;
\end{equation*}
consequently it is a continuous solution of~\eqref{mainEq}. Since $\partial_r h$ exists and is continuous and $Dh=G(h-\bar h)$ is continuous as well, $h$ is in fact a ${\mathcal C}^1$ solution. One can then show that the solution is as regular as the initial data.

\begin{Remark}
An unfortunately common misconception comes from assuming that the uniform control over the sequence $\partial_rh_n$, provided by~\eqref{bola}, is enough to conclude that  $h$ is differentiable with respect to $r$ and with that avoid the  Arzel\`a-Ascoli argument just sketched.  To see that this is not the case consider the sequence $f_n(x)$ obtained by appropriately smoothing out the corners of
$\hat f_n(x)=|x|$, if $|x|\geq 1/n$ and $\hat f_n(x)\equiv 1/n$ if $|x|<1/n$. Now, although the sequence of derivatives  $f'_n(x)$ has supremum equal to $1$ the sequence $f_n$ converges uniformly to the non-differentiable function $f(x)=|x|$.
\end{Remark}

The argument sketched here allowed Christodoulou to prove the existence of a unique global (in $u$ and $r$) solution $h\in X'$ of~\eqref{mainEq}, given appropriately small
initial data. The desired decay of the solution is already built in from the fact that the equation was solved in $X'$, and the remaining conclusions, that the respective (Bondi-spherically symmetric) space-time, determined from~\eqref{metricQuoficients}, is future geodesically complete and has vanishing final Bondi mass, follow easily from these decay properties.

\section{The positive cosmological constant case.}

Two lessons from the linear theory immediately show that the inclusion of a positive cosmological constant requires a deflection from the strategy developed for the asymptotically flat case:

First, solutions to the wave equation in a fixed de Sitter background (see equation~\eqref{wavedeSitter} for spherically symmetric waves), converge exponentially in  Bondi  time and polynomially in $r$ to a function in $\mathscr{I^+}$~\cite{Rendall:2004,CostaSpherical}; this shows the inadequacy of $Y'$ and $X'$, defined in the previous section, for the $\Lambda>0$ case. We stress that the problem is not so much the qualitative difference in decay, which in principle could  be easily accommodated and might even be occasionally beneficial, but the fact that $h$ does not decay to zero  at infinity ($r=\infty$).

Secondly, in the previous section we saw that a global solution for the zero cosmological constant case was obtained in~\cite{Christodoulou:1986} by constructing a sequence of functions which was a contraction in $Y'$; recall that the elements of $Y'$ are defined in the full domain $0\leq u,r <\infty$. Such direct strategy does not work (at least for analogous choices of function spaces) when a positive cosmological constant is considered, since a contraction in the full domain is no longer available even in the linear case. To solve the linear problem, in this framework, it was  already necessary~\cite{CostaSpherical} to construct solutions in  rectangles $[0,U]\times[0,R]$ and then find ways to extend the result to the full domain. This led us to consider  a similar strategy for the nonlinear problem~\cite{CostaProblem}.

\subsection{Norms and basic estimates}


\newcommand{\our}{\mathcal{C}^{0}_{U,R}}

\newcommand{\Xour}{X_{U,R}}


The previous discussion suggests the search of solutions in the following function spaces:
Given $U,R>0$, let $\mathcal{C}^{0}_{U,R}$ denote the Banach space
$\left(\mathcal{C}^{0}([0,U]\times[0,R]),\|\cdot\|_{\mathcal{C}^{0}_{U,R}}\right)$, where
\begin{equation*}
 \left\|h\right\|_{\mathcal{C}^{0}_{U,R}}:=\sup_{(u,r)\in[0,U]\times[0,R]}\left|h(u,r)\right|,
\end{equation*}
and let $X_{U,R}$ denote the Banach space of functions
which are continuous and have continuous partial derivative with respect to $r$, normed by

\begin{equation*}
 \|h\|_{X_{U,R}}:=\|h\|_{\our}+\|\partial_r h\|_{\our}\;.
\end{equation*}

For functions defined on $[0,R]$ we will denote $C^0([0,R])$ by $C^0_R$, $C^1([0,R])$ by $X_R$, and will also use these notations for the corresponding norms.

\vspace{0.2cm}

Our basic estimates now read
\begin{equation}
\label{barhLambda}
\|\bar h\|_{\our}\leq\| h\|_{\our}\;,
\end{equation}
\begin{equation}
\label{h-barhLambda}
\left|h(u,r)-\bar{h}(u,r)\right|\leq \frac{r}{2} \|\partial_rh\|_{\our}\;,
\end{equation}
from which
\begin{equation}
\label{fLambda}
f_{de Sitter}\equiv1\leq f\leq K:=\exp\left({C\|\partial_rh\|_{\our}^2R^2}\right)\;.
\end{equation}
For the zeroth order term we now obtain, for small $\|\partial_rh\|_{\our}$,
\begin{equation}
\label{GLambda}G\leq-C_1\, r\quad , \quad C_1=\frac{\Lambda}{3}+O(\|\partial_rh\|_{\our})\;.
\end{equation}

Some comments concerning the comparison with the estimates~\eqref{h-barhLambda0}--\eqref{GLambda0} presented for the $\Lambda=0$ case are in order:

The lack of decay in~\eqref{barhLambda} and~\eqref{GLambda} are intrinsic to the problem. In fact, we are obtaining at least linear radial growth for $|G|$.  Another intrinsic difficulty, created by the introduction of a positive $\Lambda$, is the fact that $G$ does not become small for a vanishing scalar field; recall that for $\Lambda=0$ it vanished at a cubic rate with the size of the scalar field. Nonetheless, although  the behavior of $G$, as estimated by~\eqref{GLambda},  creates difficulties  at some points, its sign  will prove extremely helpful, and its radial linear growth will be the driving force behind the exponential decay.

On the other hand, the problems present in~\eqref{fLambda}~\footnote{One should also note that when comparing these estimates with~\eqref{fLambda0}, their $\Lambda=0$ analogs, they appear reversed as a consequence of the already mentioned difference in the gauge fixing of $u$.} are technical and one should be able to overcome them in the near future. We are off course referring to the  undesirable exponential blow up, in $R$, of the upper bound for $f$. To obtain a uniform estimate, like the one for $\Lambda=0$~\eqref{fLambda0}, all is needed is  radial decay for~\eqref{h-barhLambda} of any order $r^{-\alpha}$, $\alpha>0$, which in turn follows if one imposes sufficient radial decay for the $r$--derivative of $h$, i.e.,  if one considers instead of $\Xour$ the space associated to a norm of the form:
\begin{equation}
\label{normDecay}
 \|h\|_{X_{U,R,p}}:=\|h\|_{\our}+\|(1+r)^p\partial_r h\|_{\our}\;;
\end{equation}
any choice of the form  $p\geq2-\delta$, with $0\leq\delta<1$, seems to do the trick. Unfortunately, for such norms we were not able to get the iteration scheme~\eqref{iterationII} working; more precisely we weren't even able to obtain a result analogous to~\eqref{bola}.
%
%
%

Since~\eqref{fLambda} is fundamental, as it measures the deviation from de Sitter, the constant $K$ will propagate throughout most estimates and with it its  corresponding radial exponential blow up; this  forced us to truncate our initial data and consider solutions defined only up to $r=R$, for any $R> \sqrt{3/\Lambda}$. Such restriction creates problems of its own, that will be discussed shortly, but it turns out that these ones we were able to overcome.

Note as well that estimates~\eqref{h-barhLambda}--\eqref{GLambda} depend only on the size of the radial derivative of $h$. This will be explored to establish exponential decay and  seems to open the door to potential generalizations of the main results presented here (see Section~\ref{SectionGlobalRevisited}).

\subsection{The apparent cosmological horizon and Bondi mass}
\label{SectionMass}
In~\cite{CostaProblem} it is shown that the locus $\{\tilde f=0\}$ might be parameterized by a curve $u\mapsto r_c(u)$.
Since this defines precisely the set of points where $\frac{\partial}{\partial u}$ is null, the curve $r=r_c(u)$ determines an {\em apparent cosmological horizon}. It turns out that such apparent horizon is causal and lies outside the cosmological region; such properties should be contrasted with the properties of ``typical" apparent horizons  which are acronal and lie inside the (closure of the) black hole region.

Consider the renormalized Hawking mass function
\begin{equation}
m(u,r)=\frac{r}{2}\left(1-\frac{\tilde{f}}{f}-\frac{\Lambda}{3}r^{2}\right),
\label{massfunction}
\end{equation}
which measures the mass contained within the sphere of radius $r$ at retarded time $u$, renormalized so as to remove the contribution of the cosmological constant and make it coincide with the mass parameter in the case of the Schwarzschild-de Sitter spacetime.
One can show that
$m(u,r_{c}(u))$ is a nonincreasing function of $u$. Therefore the limit
$$
M_1:=\lim_{u\to U}m(u,r_{c}(u))=\frac{r_{1}}{2}\left(1-\frac{\Lambda}{3}r^{2}_{1}\right)
$$
exists and can be seen to satisfy $0 \leq M_{1}<1/\sqrt{9\Lambda}$. We call this limit the {\em final Bondi mass}. Note that, unlike the usual definition in the asymptotically flat case, where the limit is taken at $r=+\infty$, here we take the limit along the apparent cosmological horizon. Although the original motivation for such construction came from the undesirable restriction in the radial range, the constructions itself is very natural and should be taken into account even if the radial restriction is lifted.

\subsection{Characteristics}
\label{sectionCharact}

The integral curves of $D$ through $(u_1,r_1)$ are denoted by  $\chi(u)=\chi(u;u_1,r_1)=(u,r(u;u_1,r_1))$ as before; they are the characteristics of the problem and satisfy the ordinary differential equation
\begin{equation}
 \frac{dr}{du}=-\frac{1}{2}\tilde{f}(u,r)\;.
\label{Characteristic_ODE}
\end{equation}
They are light rays and, in de Sitter, as a simple analysis of its Penrose diagram (figure~\ref{Penrose}) reveals, they separate into 3 different types: the ones converging to the center $r=0$, the ones converging to infinity $r=\infty$ and the one defining the cosmological horizon $r=\sqrt{3/\Lambda}$. We will refer to them as incoming null rays; from the previous characterization this is clearly an abuse of terminology that nonetheless seems appropriated  since the other family of null rays are all outgoing to $r=\infty$.

Estimate~\eqref{fLambda} allows us, using~\eqref{metricQuoficients}, to estimate $\tilde f$, which in turn allows us to control~\eqref{Characteristic_ODE}. By introducing
\begin{equation}
\label{rcpm}
\quad r^{-}_{c}=\sqrt{\frac{3}{\Lambda {\myK}}} \quad \text{ and } \quad  \quad r^{+}_{c}=\sqrt{\frac{3{\myK}}{\Lambda }}\;,
\end{equation}
we  obtain the following classification for the characteristic through $(u_1,r_1)$:
\begin{itemize}
\item {\bf Local region} ($r_1<r^{-}_{c}$):
\end{itemize}
\begin{equation}
 \frac{1}{2\alpha}\tanh{\left\{\alpha(c^{-}-u)\right\}}\leq r(u)\leq\frac{{\myK}}{2\alpha}\tanh{\left\{\alpha(c^{+}-u)\right\}}\;\;\;,\;\; \forall u\leq u_1\;.
\label{charLoc}
\end{equation}
\begin{itemize}
 \item {\bf Intermediate region} ($r^{-}_{c}\leq r_1< r^{+}_{c}$):
\end{itemize}
\begin{equation}
\frac{1}{2\alpha}\coth{\left\{\alpha(c^{-}-u)\right\}}\leq r(u)\leq\frac{{\myK}}{2\alpha}\tanh{\left\{\alpha(c^{+}-u)\right\}}\;\;\;,\;\; \forall
u\leq u_1\;.
\label{charInt}
\end{equation}
\begin{itemize}
 \item {\bf Cosmological region} ($r\geq r^{+}_{c}$):
\end{itemize}
\begin{equation}
 \frac{1}{2\alpha}\coth{\left\{\alpha(c^{-}-u)\right\}}\leq r(u)\leq\frac{{\myK}}{2\alpha}\coth{\left\{\alpha(c^{+}-u)\right\}}\;\;\;,\;\; \forall u\leq u_1\;.
 \label{charCosm}
\end{equation}
The exact value of the remaining constants is not of great relevance for us here, but it is important to note that $r_c^-\leq\sqrt{3/\Lambda}\leq r_c^+$ and that
\begin{equation}
\label{cosmHorizon}r_c^{\pm}\rightarrow \sqrt{3/\Lambda}\quad \text{, as } \quad \|\partial_rh\|_{\our}\rightarrow 0\;;
\end{equation}
recall that such limit is the value of the radius  of de Sitter's cosmological horizon.

\begin{figure}[h!]

\begin{center}

\psfrag{u}{$u$}

\psfrag{r}{$r$}

\psfrag{rc+}{$r_c^+$}

\psfrag{rc-}{$r_c^-$}

\psfrag{(u,r)}{$(u_1,r_1)$}

\epsfxsize=.6\textwidth

\leavevmode

\epsfbox{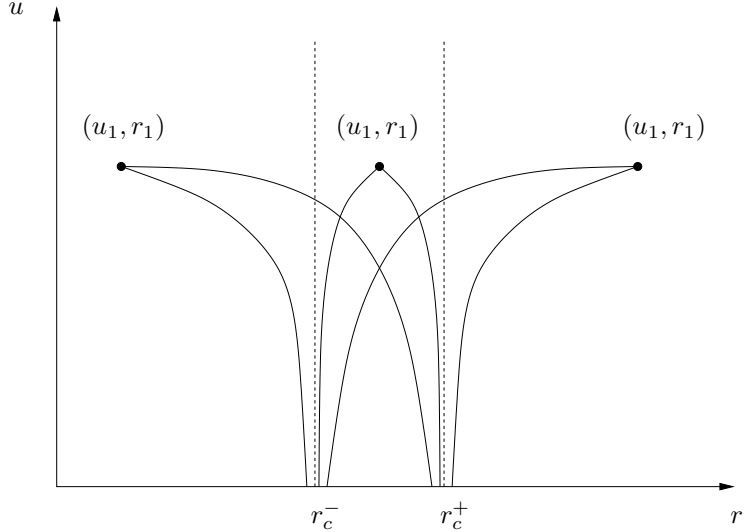}

\end{center}

\caption{Bounds for the characteristics through the point $(u_1,r_1)$ in the local ($r_1 < r_c^-$), intermediate ($r_c^- \leq r_1 < r_c^+$) and cosmological ($r_1 \geq r_c^+$) regions.} \label{Charact}

\end{figure}

A consequence of the previous classification, which will be vital in establishing exponential decay, is that: if $r(u_1)=r_1< r_c^-$, $r(u)\rightarrow 0$, in finite time, and if $r_1\geq r_c^-$ we get $r(u)\geq r_c^-$, for all $u$; this together with~\eqref{GLambda} provides, for appropriately small $\|\partial_rh\|_{\our}$, the uniform estimate
\begin{equation}
\label{eIntG}
e^{\int_u^{u_1} G\circ\chi(s)ds}\leq C e^{-C_1r_c^- (u_1-u)}\;.
\end{equation}

Another relevant consequence of the same facts, which will be central in establishing global existence,  is that, for small $\|\partial_rh\|_{\our}$ and $u_1\leq U$,
\begin{equation}
\label{C*}
\int_0^{u_1}e^{\int_u^{u_1}2\,G\circ\chi(s)ds}du\leq \int_0^{u_1}e^{-2\,C_1\int_u^{u_1}r(s)ds}du\leq C_3\quad\text{ (not depending on $U$)}\;.
\end{equation}
%

\subsection{Iteration process and local existence}
\label{SectionIteration}

After fixing $\Lambda,U>0$ and $R>\sqrt{3/\Lambda}$, and given appropriate initial data $h_0$, we consider the sequence $\{h_n\}_{n\in\mathbb{N}_0}$
defined by $h_0(u,r)=h_0(r)$ and
\begin{equation}
\label{iterationII}
 \left\{
\begin{array}{l}
 D_{n}h_{n+1}-G_{n}h_{n+1}=-G_{n}\bar{h}_{n}\\
 h_{n+1}(0,r)=h_{0}(r)\;.
\end{array}
\right.
\end{equation}
Recall, see~\eqref{iterationIntegral},  that integration along the characteristics leads to
\begin{equation}
\label{iterationIntegral2}
h_{n+1}(u_1,r_1)=h_{0}(r_n(0))e^{\int^{u_1}_{0}G_{n|_{\chi_{n}}}dv}
-\int^{u_1}_{0}\left(G_{n}\bar{h}_{n}\right)_{|_{\chi_{n}}}e^{\int^{u_1}_{u}G_{n|_{\chi_{n}}}dv}du\;\;,
\end{equation}
which we rewrite  due to its importance and in view of slight differences in the notations used in the vanishing and positive $\Lambda$ analysis.   At each step this defines a function
$$h_{n+1}:{\mathcal R}_{n+1}\subset[0,U]\times[0,R]\rightarrow \mathbb{R}\;,$$
where ${\mathcal R}_{n+1}$ is the subset of the rectangle domain foliated by the characteristics emanating from $[0,R]$, i.e.,
$${\mathcal R}_{n+1}=\{(u,r)\in[0,U]\times[0,R] \;|\; \chi_n(u)=(u,r_n(u))=(u,r)\text{ and }r_n(0)\in[0,R]\}\;.$$
A necessary condition to obtain a well defined sequence in $\Xour$ is that: ${\mathcal R}_{n+1}=[0,U]\times[0,R]$, for all $n$. By examining figure~\ref{Charact} we see that this desired property follows if $r_c^+<R$, since the characteristics that enter the region $r>r_c^+$ have an  increasing radial coordinate from then on.
According to~\eqref{cosmHorizon} this
is always the case for $R>\sqrt{3/\Lambda}$, given an appropriately small scalar field. This has to be done uniformly in $n$ and, in fact, if by induction hypothesis one has
$$ \|h_{n}\|_{X_{U,R}} \leq (1+C^*)\|h_0\|_{X_R}\;,$$
then, recalling the definition of $K$~\eqref{fLambda}, we see that
$$K_n \leq e^{C(1+C^*)^2\|h_0\|_{X_R}^2R^2}\;,$$
which, for appropriately small $\|h_0\|_{X_R}$, leads to
$$r^+_{c,n}=\sqrt{\frac{K_n\Lambda}{3}}<R\;,$$
as desired. This explains why there is no need to impose boundary conditions at $r=R$.

The core of the analysis corresponds to the following induction step:
$$\left\{\begin{array}{lc}
G_{n}\leq 0 & (H_1) \\
\|h_{n}\|_{\our}=\|h_{0}\|_{\our} & (H_2) \\
\|h_{n}\|_{\Xour}\leq (1+C^*)\|h_{0}\|_{\Xour} & (H_3) \\
\end{array}\right.
\Rightarrow
\left\{\begin{array}{lc}
G_{n+1}\leq 0 & (C_1) \\
\|h_{n+1}\|_{\our}=\|h_{0}\|_{\our} & (C_2) \\
\|h_{n+1}\|_{\Xour}\leq (1+C^*)\|h_{0}\|_{\Xour} & (C_3) \\
\end{array}\right.\;,
$$
for $C^*>0$ not depending on either $n$ or $U$ (recall~\eqref{C*}).  This is done by following the steps:
\begin{enumerate}
\item[i)] $(H_1)\wedge (H_2)\Rightarrow (C_2)$\;,
\item[ii)] $(H_1)\wedge(H_2)\wedge(C2)\Rightarrow (C_3)$\;,
\item[iii)] $(C_3)\Rightarrow (C_1)$\;.
\end{enumerate}
The first step corresponds to the following simple and fortunate observation:
\begin{equation*}
\begin{aligned}
|h_{n+1}(u_1,r_1)|
&\leq
\|h_0\|_{\mathcal{C}^0_R}\;e^{\int_0^{u_1} G_n|_{\chi_n} dv}
+\|\bar{h}_{n}\|_{\our}\; \int_0^{u_1} -G_n|_{\chi_n} e^{\int_u^{u_1} G_n|_{\chi_n}dv} du
\\
&\leq \|h_{0}\|_{\mathcal{C}^{0}_R} \underbrace{\left(e^{\int_0^{u_1} G_n|_{\chi_n} dv}
- \int_0^{u_1} G_n|_{\chi_n}  e^{\int_u^{u_1}G_n|_{\chi_n}  dv} dv\right)}_{\equiv1} =\|h_{0}\|_{\mathcal{C}^{0}_R}\; .
\end{aligned}
\end{equation*}
To derive step ii) one obtains an evolution equation for $\partial_rh_n$ by differentiating~\eqref{iterationII}. When integrated along the characteristics this equations yields
\begin{equation}
\label{drhn}
 \begin{aligned}
\partial_r h_{n+1}(u_{1},r_{1})
 &=
 \partial_r h_{0}(\chi_n(0))\,e^{\int^{u_{1}}_{0}2 G_{n|_{\chi_{n}}}dv}
 \\
 &-
 \int^{u_1}_{0}\left[{J_{n}\partial_r\bar{h}_{n}}+\underbrace{\left(J_{n}-G_{n}\right)\frac{(h_{n+1}-h_{n})}{r}}_{\dagger}\right]_{|_{\chi_{n}}}e^{\int^{u_1}_{u}2 G_{n|_{\chi_{n}}}dv}du\;,
\end{aligned}
\end{equation}
with, recall~\eqref{defJ},
\begin{equation}
\label{J}
|J_n|\leq C_2\, r\quad , \quad C_2=O(\|\partial_rh\|_{\our})\;.
\end{equation}
Relying also on~\eqref{C*}, one can then obtain conclusion $(C_3)$.

One can now  hopefully get a glimpse on some of the reasons why is not so easy  to extend the previous analysis to the case with decay, i.e., when considering the norms~\eqref{normDecay} with $p>0$: recall that we cannot expect $h_n$ nor $G_n$ to decay, and so we run into trouble in trying to get the term $\dagger$ in~\eqref{drhn} to decay; also, since (for $\Lambda>0$) $G_n$ does not vanish with the scalar field, when estimating $\|h\|_{X_{U,R,p}}$, the contribution from $\dagger$ seems to lead to an expression which is linear on the size of the scalar field (to see why this is a problem recall~\eqref{bola2} and the discussion following it).
Adding decay therefore requires some new ideas (see also Section~\eqref{SectionWhat}).
In the case without decay everything works out in view of the remarkable $(H_2)$ and of $(C_3)$.

We are left with step iii) which requires only~\eqref{GLambda}.

Using the established estimates $(C_1)$--$(C_3)$, which hold for all $n$, we can now show that, for small initial data and small enough $U$, the sequence $h_n$ contracts with respect to the $\|\cdot\|_{\Xour}$ norm, i.e.,
$$\|h_{n+1}-h_n\|_{\Xour}\leq \sigma \|h_{n}-h_{n-1}\|_{\Xour}\;,$$
for some $\sigma<1$. Since this norm also includes the $r$--derivative, the resulting limiting function will be differentiable with respect to $r$, and consequently we obtain a $C^1$ solution.  We were able to avoid the all Arzel\`a-Ascoli argument used for the asymptotically flat case, see Section~\eqref{SectionLambda0}, because, in essence, we are now asking for less by allowing ourselves to make $U$ small.

In fact, one can prove the following~\cite{CostaProblem}:

\begin{thm}
\label{thm1}
Let $\Lambda>0$, $R>\sqrt{\frac{3}{\Lambda}}$ and
 $h_{0}\in\mathcal{C}^{k}([0,R])$ for $k \geq 1$.
For appropriately small  $\|h_{0}\|_{X_{R}}$,  the initial value problem
\begin{equation}
\label{mainEq0}
 \left\{
\begin{array}{l}
  Dh = G\left(h-\bar{h}\right) \\
  h(0,r)= h_0(r)\;,
\end{array}
\right.
\end{equation}
has a unique solution $h\in\mathcal{C}^{k}([0,U]\times[0,R])$, for $U=U(\|h_{0}\|_{X_{R}};R,\Lambda)$ sufficiently small. Moreover,
\begin{equation}
\label{boundh1}
\left\|h\right\|_{\mathcal{C}^{0}_{U,R}}=\left\|h_{0}\right\|_{\mathcal{C}^{0}_{R}}
\end{equation}
and
\begin{equation}
\label{boundh2}
\left\|h\right\|_{X_{U,R}}\leq (1+C^*)\,\left\|h_{0}\right\|_{X_{R}} \quad \text{, $C^*$ not depending on $U$}.
\end{equation}
\end{thm}

 \begin{Remark}
 Note that  the existence time $U\rightarrow 0$, as $R\rightarrow \infty$. This undesirable feature can be traced back to~\eqref{fLambda}.
 \end{Remark}

This is an unusual statement since it requires both the existence time $U$ and the size of the data to be small. In fact we expect that the existence and uniqueness claims presented here should follow for small $U$ and arbitrarily large initial data; but note that for large data one expects to lose the estimates~\eqref{boundh1} and~\eqref{boundh2}.  We did not try to establish such result in~\cite{CostaProblem} since we were mainly interested in global existence and, in fact, in view of~\eqref{boundh1} and~\eqref{boundh2}, the local existence result provided by Theorem~\ref{thm1} is in essence a global in (Bondi) time result as we will see in the following section.

\subsection{Global existence in Bondi time}
\label{sectionGlobal}

Let us now sketch the way to construct a global (in time) solution $h:[0,\infty[\times[0,R]\rightarrow \mathbb{R}$ starting from the local solution provided by Theorem~\ref{thm1}:

Given appropriately small initial data, one starts by solving  the local (in $u$) problem and obtain a solution
\begin{equation*}
h_1:[0,U_1]\times[0,R]\rightarrow \mathbb{R}\quad,\quad U_1=U(\Lambda,R,\|h_0\|_{X_R})>0\;,
\end{equation*}
for which the following estimates hold:
\begin{equation*}
\left\|h_1\right\|_{\mathcal{C}^{0}_{U_1,R}}=\left\|h_{0}\right\|_{\mathcal{C}^{0}_{R}}\;,
\end{equation*}
and
\begin{equation}
\label{estimate2}
\|h_1\|_{X_{U_1,R}}\leq (1+C^*)\|h_0\|_{X_R}\;.
\end{equation}
We then solve once more the problem now taking as  initial data $h_1(U_1,\cdot)$; this is possible since according to~\eqref{estimate2}, by shrinking the size of the initial data if necessary, $\|h_1(U_1,\cdot)\|_{X_R}$ remains appropriately small. This time one  obtains a solution
$$h_2:[0,U_2]\times[0,R]\rightarrow \mathbb{R}\quad,\quad U_2=U(\Lambda,R,\|h_1(U_1,\cdot)\|_{X_R})>0\;.$$
We can then extend the first solution by defining
\begin{equation*}
 h(u,r):=\left\{
\begin{array}{l}
  h_{1}(u,r)\quad,\quad u\in[0,U_1]  \\
  h_{2}(u,r)\quad,\quad u\in[U_1, U_{1}+U_{2}]\;.
\end{array}
\right.
\end{equation*}
It turns out that, since $C^*$ does not depend on the existence time, this solution also satisfies
$$\|h\|_{X_{U_1+U_2,R}}\leq (1+C^*)\|h_0\|_{X_R}\;,$$
and consequently may be extend  by the same amount $U_2$ as before. This process can be repeated indefinitely, leading to a global in time solution satisfying the bounds
\begin{equation}
\label{boundh3}
\|h\|_{{\mathcal C}^0([0,\infty)\times[0,R])} = \|h_{0}\|_{{\mathcal C}^0([0,R])}\;,
\end{equation}
and
\begin{equation}
\label{boundh4}
\|h\|_{{X}([0,\infty)\times[0,R])} \leq (1+C^*) \|h_{0}\|_{{X}([0,R])}\;.
\end{equation}

This argument is neither standard nor very flexible. In Section~\ref{SectionExponential} we will sketch an idea for a more standard argument, at the cost of losing estimate~\eqref{boundh3}, which might presumably be extended to a more general setting, possibly including some classes of nonlinear scalar fields for which~\eqref{boundh3} will be unavailable.

\subsection{Exponential decay}
\label{SectionExponential}

Given a global solution $h$ as constructed in the previous section, exponential decay follows from energy estimates for
\begin{equation}
\label{energy1Def}
{\mathcal E}_1(u):=\|\partial_rh(u,\cdot)\|_{{\mathcal C}^0_R}\;.
\end{equation}
One motivation to consider this definition of energy comes from the linear theory:  although the spherically symmetric wave equation on a de Sitter background~\eqref{wavedeSitter} is not trivial, as opposed to what happens in the asymptotically flat case, from it a trivial evolution equation for $\partial_rh$ follows; this provides a mechanism for exponential decay in the uncoupled case~\cite{CostaSpherical}.

Since $\partial_rh$ satisfies equation~\eqref{D_partial_h}, integrating along the characteristics from $u=u_0$ one can show that, for small data, using~\eqref{boundh4} and recalling~\eqref{eIntG},
$${\mathcal E}_1(u_1)\leq C {\mathcal E}_1(u_0)e^{-\hat Cu_1}+\tilde{C}R\int_{u_0}^{u_1} {\mathcal E}_1(u)e^{-\hat{C}(u-u_1)}du\;. $$
Then, applying Gronwall's Lemma to ${\mathcal F}_1(u)=e^{\hat Cu}{\mathcal E}_1(u)$, we see that
$${\mathcal E}_1(u_1)\leq C {\mathcal E}_1(u_0) \exp((\tilde C-\hat C)u_1)\;.$$
It turns out that $\tilde C\rightarrow 0$ while $\hat C$ remains positive in the limit when the norm of the radial derivative of the scalar field vanishes, which can be traced back to estimates~\eqref{J} and~\eqref{GLambda}.  Using these facts one eventually obtains
\begin{equation}
\label{energyEst0}
{\mathcal E}_1(u_1)\leq C{\mathcal E}_1(0)e^{-2\hat H(u_0)u_1}\;,
\end{equation}
with $\hat H(u_0)>0$. A remarkable fact is that the previous exponential decay results in the exponential decay of $\hat H$ to the desired rate $H=\sqrt{\Lambda/3}$ a fact that can be exploited to obtain such (expected to be sharp)  decay rate; one uses the freedom to solve the problem with arbitrary starting time $u=u_0$ and then take it to be large enough to exploit the damping effects of the exponential decay and consider problems with smaller  initial data.

\subsection{Global existence revisited}
\label{SectionGlobalRevisited}

We will now provide the ideas for an alternative, more standard and flexible, argument to prove global existence based on energy estimates.  We will start by assuming a local in time existence and uniqueness result that for arbitrary initial data in $X_R$ provides a solution to~\eqref{mainEq0} with existence time
$$U=U(\|h_0\|_{X_R};\Lambda,R)>0\;.$$
To our knowledge such result has not been proven anywhere but is widely expected to be true.~\footnote{In fact all that is needed to carry out the  argument to follow is a somewhat less orthodox local existence result for initial data with (arbitrary) finite $\|h_0\|_{{\mathcal C}_R}$ and small $\|\partial_rh_0\|_{{\mathcal C}_R}$. Such result has the advantage of following from the arguments in~\cite{CostaProblem} leading to the local existence result there.}

Now given a solution $h$ to~\eqref{mainEq0} let
$U^*\geq U(\|h_0\|_{X_R})>0$
be its maximal existence time, meaning that any other solution $\hat h:[0,\hat U]\times [0,R]\rightarrow\mathbb{R}$
to~\eqref{mainEq0}, with the same initial data, will necessarily have existence time $\hat U\leq U^*$ and will coincide with $h$ on $[0,\hat U]\times [0,R]$.

Define
\begin{equation}
\label{energy1Def}
{\mathcal E}_0(u):=\|h(u,\cdot)\|_{{\mathcal C}^0_R}\;.
\end{equation}
%

\newcommand{\mE}{{\mathcal E}}


From the (assumed) local existence result the following extension principle follows:
\begin{center}
either $U^*=\infty$  or  $({\mathcal E}_0+{\mathcal E}_1)(u)$ is unbounded in $[0,U^*)$\;,
\end{center}
with ${\mathcal E}_1$ as defined in the previous section.~\footnote{If one considers instead a local existence theorem for finite $\|h_0\|_{{\mathcal C}_R}$ and small $\|\partial_rh_0\|_{{\mathcal C}_R}$ the previous extension principle has to be reformulated accordingly: for a given  $\bar \mE_1>0$, then $U^*=\infty$  or   ${\mathcal E}_0$ is unbounded in $[0,U^*)$  or $\sup_{u\in[0,U^*)}{\mathcal E}_1(u)>\bar \mE_1$.    }

We would like to recycle the argument of the previous section to control ${\mathcal E}_1$. This can in fact be done but some care has to be taken since now we do not have~\eqref{boundh4} which gave the necessary a priori control over the zeroth order term $G$ and over the characteristics; without this a priori bound all we know is that such elements are controlled by ${\mathcal E}_1$, so we need to bootstrap:


Let $\mE^*>0$ be such that if $\mE_1(u)\leq \mE^*$, for all $u\leq U$, then~\eqref{eIntG} holds. Define
$${\mathcal U}=\{U\in[0,U^*)\,:\, \mE_1(u)\leq \mE^*, \text{ for all } 0\leq u\leq U\}\;.$$
For sufficiently small initial data it is clear, by continuity, that ${\mathcal U}\neq \emptyset$. From the previous section, see~\eqref{energyEst0}, given $U\in{\mathcal U}$ one has
\begin{equation}
\label{energyEst}
{\mathcal E}_1(u)\leq C{\mathcal E}_1(0)\;,
\end{equation}
for all $u\leq U$, with $C>0$ not depending on $U$. Fix $\epsilon >0$ and consider initial data satisfying $\mE_1(0)\leq \mE^*/(C+\epsilon)$, with $C$ the constant in~\eqref{energyEst} so that the upper bound for the initial data does depend on $U$; from the above we then conclude that $\mE_1(u)\leq \mE^*$, for all $u\leq U_1$, with $U_1>U$. This shows that ${\mathcal U}$ is open in $[0,U^*)$ and since it is clearly closed as well we conclude that, for appropriately small initial data, ${\mathcal U}=[0,U^*)$.  In particular~\eqref{energyEst} holds in $[0,U^*)$ which provides a substitute for the originally missing estimate~\eqref{boundh4} and allows us to carry out the argument in the previous section leading to exponential decay for $\mE_1$.

Using the previously formulated extension principle, we see that, for small data, potential obstructions to global existence have to come from the blow up of ${\mathcal E}_0$ in finite time. To control this term note that integrating~\eqref{mainEq0} along the characteristics leads to
\begin{equation}
{\mathcal E}_0(u_1)\leq{\mathcal E}_0(0)\,e^{\int_0^{u_1} G\circ\chi(u)du}+\int_{0}^{u_1}{\mathcal E}_0(u)G\circ\chi(u)e^{\int_{u}^{u_1}G\circ\chi(s)ds}du\,.
\end{equation}
Now, since $G$ and the characteristics are controlled by the size of $\partial_rh$, they are, according to~\eqref{energyEst} controlled by ${\mathcal E}_1(0)$. Consequently the techniques used in the previous section to estimate ${\mathcal E}_1$ may be used once more to obtain
$${\mathcal E}_0(u_1)\leq C {\mathcal E}_0(0) \exp((\tilde C-\hat C)u_1)\;,$$
for new constants also depending on ${\mathcal E}_1(0)$. It turns out that these constant satisfy $\tilde C-\hat C\rightarrow \check C>0$ when
${\mathcal E}_1(0)\rightarrow 0$, a fact that can be traced back to~\eqref{GLambda} and~\eqref{J}. It will then follow that, for appropriately small initial data,
\begin{equation}
\label{energyEst2}
{\mathcal E}_0(u_1)\leq \check C_1 e^{\check C_2 u_1}\;,
\end{equation}
with $\check C_2>0$ depending on ${\mathcal E}_1(0)$ and $\check C_1>0$ depending on both ${\mathcal E}_1(0)$ and  ${\mathcal E}_0(0)$.
This time, and as expected, the energy estimate~\eqref{energyEst2} does not provide any decay but is enough to conclude that ${\mathcal E}_0$ does not blow up in finite time.

A realization of this argument would thus provide an alternative proof of global existence. This is of interest especially since it might lead to the generalization of some of the results discussed here to other matter models, for instance, to some classes of nonlinear scalar fields. On the contrary, the argument developed in~\cite{CostaProblem} and sketched earlier  is rigidly based on estimate~\eqref{boundh1} which we do not expect to generalize to other matter models besides self gravitating scalar-fields with positive cosmological constant. Of course the original argument in~\cite{CostaProblem} is preferable in this restricted context since it provides~\eqref{boundh3} which, as seen from~\eqref{energyEst2}, is lost in the energy estimate argument.

\section{What's next}
\label{SectionWhat}

We finish this article discussing future plans of research:

\subsubsection{Small data: Global existence in the radial coordinate and nonlinear scalar fields}

The necessity to restrict our analysis to a finite radial range $r\leq R$, for any $R\geq \sqrt{3/\Lambda}$, can be traced back to~\eqref{fLambda}.
As already discussed, such undesirable exponential blow up of the upper bound can be easily bypassed by restricting the function space to functions with sufficiently fast polynomial decay in $r$~\eqref{normDecay}, but then, the iteration scheme used to establish existence becomes ill-defined, at least for the simplest choices of polynomial radial decay. Recall also that such difficulty is not present in Christodoulou's analysis of the $\Lambda=0$ case where there is great freedom in choosing a decay rate. By now, we have obtained some preliminary results, for $\Lambda>0$, by a fine-tuned choice of decay and by exploring more thoroughly the decay provided by the factors $exp(\int 2G|_{\chi})$ in~\eqref{drhn}.  We hope these will allow us to generalize the results in Section~\ref{SectionIteration} to the full radial range; this is of course a fundamental step but if it will lead to a corresponding generalization of our main results to the full radial range remains to be seen.  Observe as well that by restricting the radial range the corresponding solutions contain only ``one point" of $\mathscr{I^+}$
%
%
restricting the scope of the asymptotic analysis. On the other hand, if a global in $r$ result is obtained then a full asymptotic analysis may be carried out in the hope of extending the  asymptotic approach to de Sitter to a full neighborhood of $\mathscr{I^+}$,  and give a more complete realization of the cosmic ``no-hair" conjecture.

Another, already discussed, potential line of research is to realize and expand the ideas in Section~\eqref{SectionGlobalRevisited} in order to generalize the main results presented here to some classes of nonlinear scalar fields: for instance, the spherically symmetric context might provide a good starting points to study classes of potentials with degenerate minimums, at $\phi=0$, satisfying $V(0)>0$, $V^{(k)}(0)=0$, for $k=1,\ldots, 2m-1$, and  $V^{(2m)}(0)>0$, for some integer $m\geq2$\;, for which the exact asymptotic behavior in unknown. Note that the $m=1$ case, without symmetry restrictions, is covered by the results in~\cite{Ringstrom:2008}.

\subsubsection{Large data and the formation of cosmological black holes}

The main interest in considering spherical symmetry is that, in principle, the existing nonlinear analysis techniques allow us to study the long term behavior of the corresponding $1+1$ systems for large data as well. For large data it is expected that gravitational collapse will lead to the formation of (cosmological) black holes and, foreseing such scenario,  the formulation of the cosmic ``no-hair" conjecture should be strengthened by substituting de Sitter by Schwarzschild-de Sitter in the role of attractor.  Schwarzschild-de Sitter is in fact, for fixed positive cosmological constant,  a  one-parameter (the mass)  family of solutions of the Einstein vacuum equations with positive cosmological constant; for a given range of positive values of the mass parameter, the maximal analytic extension of the corresponding solution corresponds to a space-time containing an infinite number of causally unrelated black holes and cosmological regions.  Note also that such strengthening of the statement of the cosmic ``no-hair" conjecture is not in conflict with the original one: first of all, de Sitter corresponds to the member of the Schwarzschild-de Sitter family with vanishing mass parameter and, moreover, all the solutions of the Schwarzschild-de Sitter family approach de Sitter asymptotically, for large enough radius, within their cosmological regions.

We foresee two possibilities to attack the large data case. The first would be to follow Christodoulou's construction~\cite{ChristodoulouGlobal} of weak solutions of~\eqref{mainEq} and then try to extend the deep analysis of the causal structure and asymptotic behavior of such solutions carried out in~\cite{ChristodoulouStructure, ChristodoulouMathematical}. The formation of a cosmological black hole will presumably be signaled by a non vanishing final Bondi mass. As we already saw, in the small data case, the scalar field dissipates, the final Bondi mass vanishes and no black hole forms. Such approach contains two drawbacks, already present in the original analysis of the vanishing cosmological constant case: First it will be hard to extract conditions at the level of initial data that guarantee the formation of black holes. Secondly, the class of weak solutions might have enough regularity to allow to obtain long term existence of solutions for large data but not enough to prove uniqueness. The lack of uniqueness would not allow for a realization of the conjecture under discussion.


An alternative approach is to start from the  {\em  maximal globally hyperbolic development} (MGHD) for an appropriate Cauchy problem, for instance as  provided  by~\cite{RingstromCauchy}, and then study its global structure. One shouldn't confuse the {\em maximal} terminology used here with the notion of global solution form the theory of differential equations that was used before. Once a coordinate system is fixed, for example the Bondi coordinates defined above, the   maximal  development provided by the referred generalization of the Choquet Bruhat-Geroch theorem may fail to be global in such coordinate system. It is also presumable that
well posedness of the systems that result from the field equations by gauge fixing, in principle either Bondi coordinates or double null coordinates away from the center $r=0$, must be set forth in order to study the global properties of the maximal global developments; note for instance that the presented results concerning future causal geodesical completeness and global stability do not follow from the Choquet Bruhat-Geroch theorem.

A problem to address concerns the derivation of conditions, at the level of initial data, guaranteeing the existence of trapped surfaces within the MGHD.
For the self-gravitating scalar fields, with vanishing cosmological constant, Christodoulou~\cite{ChristodoulouFormation} established the first result concerning the formation of trapped surfaces provided that the dimensionless size of an annular region of an initial cone is small enough and the dimensionless mass content of such region is large enough.
In general it is not clear if the presence of trapped surfaces implies the existence of a non-empty black hole region. Nonetheless, in spherical symmetry, this is an immediate consequence of the Raychaudhuri  equations.

Given the existence of a non-empty black hole region within the maximal development, one should try to establish a Price law providing exponential decay of a spherically symmetric self-gravitating scalar field with positive cosmological constant. From this, an extraordinary realization of the cosmic ``no-hair" conjecture, with Schwarzschild-de Sitter as attractor, would follow. A Price law, providing polynomial decay, was established for the $\Lambda=0$ case by Dafermos and Rodnianski~\cite{DafermosProof}. A relevant part of the analysis there rests on a priori estimates derived using the {\em red-shift effect}, a mechanism that provides fast decay along the event horizon; herein rests the importance of considering space-times with non-empty black hole regions. For positive $\Lambda$ there will also exist a cosmological horizon along which more {\em red-shift} will presumably help enforce decay.

\section*{Acknowledgements}

We thank  Simone Calogero, Piotr Chru\'sciel, Mihalis Dafermos, Marc Mars, Filipe Mena, Jos\'e Nat\'ario, Ra\"ul Vera, Alan Rendall and Hans Ringstr\"om for useful discussions and comments. Special thanks are due to J. Nat\'ario for various comments concerning a preliminary version of this work.   The author would like to acknowledge the hospitality afforded to him by the  Department of Pure Mathematics and Mathematical Statistics of the University of Cambridge, as well as by the Erwin Schr\"odinger Institute (Vienna), during the workshop ``Dynamics of General Relativity: Black Holes and Asymptotics''. This work was partially supported by project CERN/FP/116377/2010.


\begin{thebibliography}{{Chr}87b}

\bibitem[{Bey}]{Beyer:2010}
{Beyer, Florian}, \emph{{The cosmic no-hair conjecture: A study of the Nariai
  solutions}}, Proceedings of the Twelfth Marcel Grossmann Meeting on General
  Relativity, Edited by T. Damour, R.T. Jantzen, R. Ruffini.

\bibitem[{Chr}86a]{ChristodoulouGlobal}
{Christodoulou, Demetrios}, \emph{{Global Existence of Generalized Solutions of
  the Spherically Symmetric Einstein Scalar Equations in the Large}},
  Commun.Math.Phys. \textbf{106} (1986), 587--621.

\bibitem[{Chr}86b]{Christodoulou:1986}
\bysame, \emph{{The Problem of a Self-gravitating Scalar Field}}, Commun. Math.
  Phys. \textbf{105} (1986), 337--361.

\bibitem[{Chr}87a]{ChristodoulouMathematical}
\bysame, \emph{{A Mathematical Theory of Gravitational Collapse}},
  Commun.Math.Phys. \textbf{109} (1987), 613--647.

\bibitem[{Chr}87b]{ChristodoulouStructure}
\bysame, \emph{{The Structure and Uniqueness of Generalized Solutions of the
  Spherically Symmetric Einstein Scalar Equations}}, Commun.Math.Phys.
  \textbf{109} (1987), 591--611.

\bibitem[{Chr}91]{ChristodoulouFormation}
\bysame, \emph{{The formation of black holes and singularities in spherically
  symmetric gravitational collapse}}, Comm. Pure Appl. Math. \textbf{44}
  (1991), 339--373.

\bibitem[{Chr}09]{Christodoulou:2008}
\bysame, \emph{{The Formation of Black Holes in General Relativity}}, EMS
  Monographs in Mathematics. (2009), arXiv/gr-qc:0805.3880.

\bibitem[{Cos}]{CostaProblem}
{Costa, Jo\~ao L. and Alho, Artur and Nat\'ario, Jos\'e}, \emph{{The problem of
  a self-gravitating scalar field with positive cosmological constant}},
  Annales Henri Poincar\'e \textbf{To appear}.

\bibitem[{Cos}12]{CostaSpherical}
\bysame, \emph{{Spherical linear waves in de Sitter spacetime}}, J.Math.Phys.
  \textbf{53} (2012), 052501.

\bibitem[{Daf}05]{DafermosProof}
{Dafermos, Mihalis and Rodnianski, Igor}, \emph{{A Proof of Price's law for the
  collapse of a selfgravitating scalar field}}, Invent.Math. \textbf{162}
  (2005), 381--457.

\bibitem[HS12]{HolzegelSelf}
Gustav Holzegel and Jacques Smulevici, \emph{{Self-gravitating Klein-Gordon
  fields in asymptotically Anti-de-Sitter spacetimes}}, Annales Henri Poincare
  \textbf{13} (2012), 991--1038.

\bibitem[HS13]{HolzegelStability}
\bysame, \emph{{Stability of Schwarzschild-AdS for the spherically symmetric
  Einstein-Klein-Gordon system}}, Commun.Math.Phys. \textbf{317} (2013),
  205--251.

\bibitem[{Ren}04]{Rendall:2004}
{Rendall, Alan D.}, \emph{{Fuchsian methods and space-time singularities}},
  Class.Quant.Grav. \textbf{21} (2004), S295--S304.

\bibitem[{Rin}08]{Ringstrom:2008}
{Ringstr\"om, Hans}, \emph{{Future stability of the Einstein-non-linear scalar
  field system}}, Invent. Math. \textbf{173} (2008), 123–208.

\bibitem[{Rin}09]{RingstromCauchy}
\bysame, \emph{{The Cauchy Problem in General Relativity}}, Lectures in
  Mathematics and Physics, European Mathematical Society (2009).

\end{thebibliography}

\providecommand{\bysame}{\leavevmode\hbox to3em{\hrulefill}\thinspace}
\providecommand{\MR}{\relax\ifhmode\unskip\space\fi MR }
\providecommand{\MRhref}[2]{%
  \href{http://www.ams.org/mathscinet-getitem?mr=#1}{#2}
}
\providecommand{\href}[2]{#2}

\end{document}